# UV/Near-IR dual band photodetector based on p-GaN/α-In$_2$Se$_3$ heterojunction

*Swanand V. Solanke[1*], Rohith Soman[1,2], Muralidharan Rangarajan[1], Srinivasan Raghavan[1] and Digbijoy N. Nath[1]*

[1]Centre For Nano Science and Engineering (CeNSE), Indian Institute of Science, Bengaluru, Karnataka -5600129 (INDIA)

[2]Depart of Electronic and Communication Engineering, Indian Institute of Science, Bengaluru, Karnataka -560012 (INDIA)

**Abstract**

In this report, we demonstrate dual band vertical heterojunction photodetector realized by integrating α-In$_2$Se$_3$ with p-type GaN. Flakes of ~ 110 nm thickness were exfoliated on MOCVD grown p-GaN on silicon substrate. Devices showed two distinct detection peaks in spectral responsivity, one at 365 nm and another at 850 nm, corresponding to band edges of GaN and α-In$_2$Se$_3$ respectively, with considerable rejection in visible spectrum. Normalised responsivity values were found out to be ~70 mA/W at both 365 nm and 850 nm for the bias of -3V along with photo-to-dark current ratio of ~665 and ~75 in that order. The Devices also showed fast transient response with no persistent photoconductivity (PPC). The specific detectivity values estimated were ~10$^{11}$ Jones and ~10$^{10}$ Jones corresponding to illumination at 365 nm and 850 nm respectively. A good linearity of ~0.4 was observed in power dependent analysis of spectral responsivity at 365 nm. The device performance, post annealing was also studied. This study is expected to pave way for new type of optoelectronic devices by integrating direct bandgap layered material like α-In$_2$Se$_3$ and wide bandgap semiconductors.

## 1. Introduction

Multiband or multi-spectral photodetectors have gained a lot of interest for their promise in enhancement of detection of target in clutter, ability to distinguish between intended targets and decoys in addition to remote temperature measurement capability, when it comes to military applications. Until now, the multispectral imagery data acquisition has been achieved by overlaying images captured in two or multiple different spectra by sensor arrays. Despite appearing to be simple, such systems require several bulky and expensive optical components in addition to high-end image processing tools for reliable image overlapping. However, in a complex or harsh environment, capacity of photodetector can be greatly enhanced by multiband photodetection in a single device, since different types of information can be gathered from different types of targets using single distinguishable spectrum response.[1][2][3] A monolithic device with multi-wavelength detection capability is highly desired in such scenario owing to its several advantages, especially in terms of robustness, fabrication cost and operational ease and flexibility.

There have been reports of multi-spectral photodetectors realized on epitaxially grown alloyed semiconductors, layered-materials (2D semiconductors), multiple quantum wells and superlattices and on different types of heterojunctions.[4][5][6] With rapid development in different kinds of semiconductor growth techniques, many sophisticated multiband photodetectors have been realized. However, when it comes to dual-band photodetection in two different spectra, say UV and IR, realization becomes increasingly difficult. The main limiting factors are availability of suitable semiconductors (mainly direct bandgap and in extreme spectral regions), growth of semiconductors with vastly different bandgaps over another without lattice mismatch, difficulties in doping and metal contact formation (especially in wide bandgap semiconductors). These difficulties arise even though two different semiconductors may belong to the same group of alloys. For example, it is extremely difficult to grow InN or InGaN with higher Indium contex (suitable for IR) over AlGaN (suitable for UV) because of severe lattice mismatch and growth challenges.

In such a scenario, layered materials (or '2D materials' as commonly called), such as transition metal dichalcogenides (TMDs) or III-VI group semiconductors attracted significant attention due to the fact that they can be easily transferred on virtually any substrate without the worry of lattice mismatch as the layers are held out-of-plane by Van-der-Waals force.[7][8] Besides, many of these layered materials have exciting electrical and optical properties. Several groups have reported layered material (LM) based heterojunctions, LM/3D heterojunctions towards multifunctional device applications, most of which emphasize on diode-like applications with vertical current transport and single band photodetection.[9][10][11] However, study of heterojunctions between layered materials and conventional 3D semiconductors for dual-band photodetection are at nascent stage, especially heterojunctions combining direct bandgap layered materials and wide bandgap semiconductors for extreme bandgap engineering toward optoelectronic devices. For example, MoS$_2$, which is one of the most widely studied layered materials, has been primarily investigated for single band photodetection.[12][13] Recently, a few direct bandgap 2D materials like In$_2$Se$_3$, GeSe (III-VI group) are now gaining traction in the photodetector community. Among III-IV group materials, In$_2$Se$_3$ has shown to be of particular interest and is reported to possess five crystalline phases (α, β, γ, δ and κ) among which α and β have direct bandgap of ~1.45 eV and ~1.3 eV respectively. It is also reported to have good sensitivity and absorption (~10$^5$ cm$^{-1}$), making it suitable for Visible-NIR optoelectronics as well as for photovoltaics.[14][15][16] The carrier mobility in In$_2$Se$_3$ has been reported to be 30 +/- 8 cm$^2$/Vs, which is comparable to multilayer MoS$_2$.[17] On the other extreme, wide band gap semiconductor such as GaN is at the forefront of technological revolution having altered the lighting landscape with the ubiquitous white LED bulbs.

Here, we report on UV-near IR dual-band photodetector by realizing as heterojunction between relatively unexplored multilayer exfoliated α-In$_2$Se$_3$ with MOCVD-grown p-type GaN. By exploiting relative band alignments between these two materials, photo generated carriers could be extracted, thus leading to simultaneous detection in both UV and near-IR regime with considerable rejection in the visible spectrum.

## 2. Device structure, material characterization and fabrication

The Mg-doped p-GaN, necessary for the transfer of α-In$_2$Se$_3$ was grown on 2-inch <111> Si wafer using Metal-Organic Chemical vapor Deposition (MOCVD) method in Aixtron 200/4-HT horizontal flow low-pressure reactor. The details of the growth are reported elsewhere.[18][19][20]  The concentration of holes was estimated to be 1.1x10$^{16}$ cm$^{-3}$ using Van-der-pauw done on a reference sample diced from the same 2 inch p-GaN wafer. As grown p-GaN surface exhibited good surface smoothness with average rms smoothness of 0.6 nm. The structural quality analysis was performed with the help of Rocking Curve (RC) scans for (002) plane and (102) plane in 4-circle XRD (Rigaku) tool. Figure S1 (supplementary data) shows the corresponding FWHM values for (002) and (102) plane, 0.245 and 0.620 degrees respectively along with RC scans.

A piece of p-GaN was then cleaned using Acetone-IPA-DI water (3 minutes each) in ultrasonic bath. This was followed by exfoliation of α-In$_2$Se$_3$ (obtained from bulk sample purchased from hqGraphene) flakes using standard scotch tape method. By Inspection through optical microscope multilayer flakes were chosen for enhanced light absorption. Although other techniques to enhance light absorption are reported,[21][22]  the use of multi-layered material ruled out the necessity to control the flake thickness along with the use of top absorbing layer or reflecting layer underneath. The multi-layer (~110 nm) nature of the flakes was confirmed by the step height measurement carried out using AFM (Bruker) (figure S2). The Raman Spectroscopy (Horiba LabRam HR) was then performed using 532 nm laser to confirm the α phase of In$_2$Se$_3$. Figure S3 shows the Raman spectra for α-In$_2$Se$_3$ wherein the $E^1$, $A^1_1$, $A^2_1$ and $E^4$ peaks can be observed. [16][23][24][25]

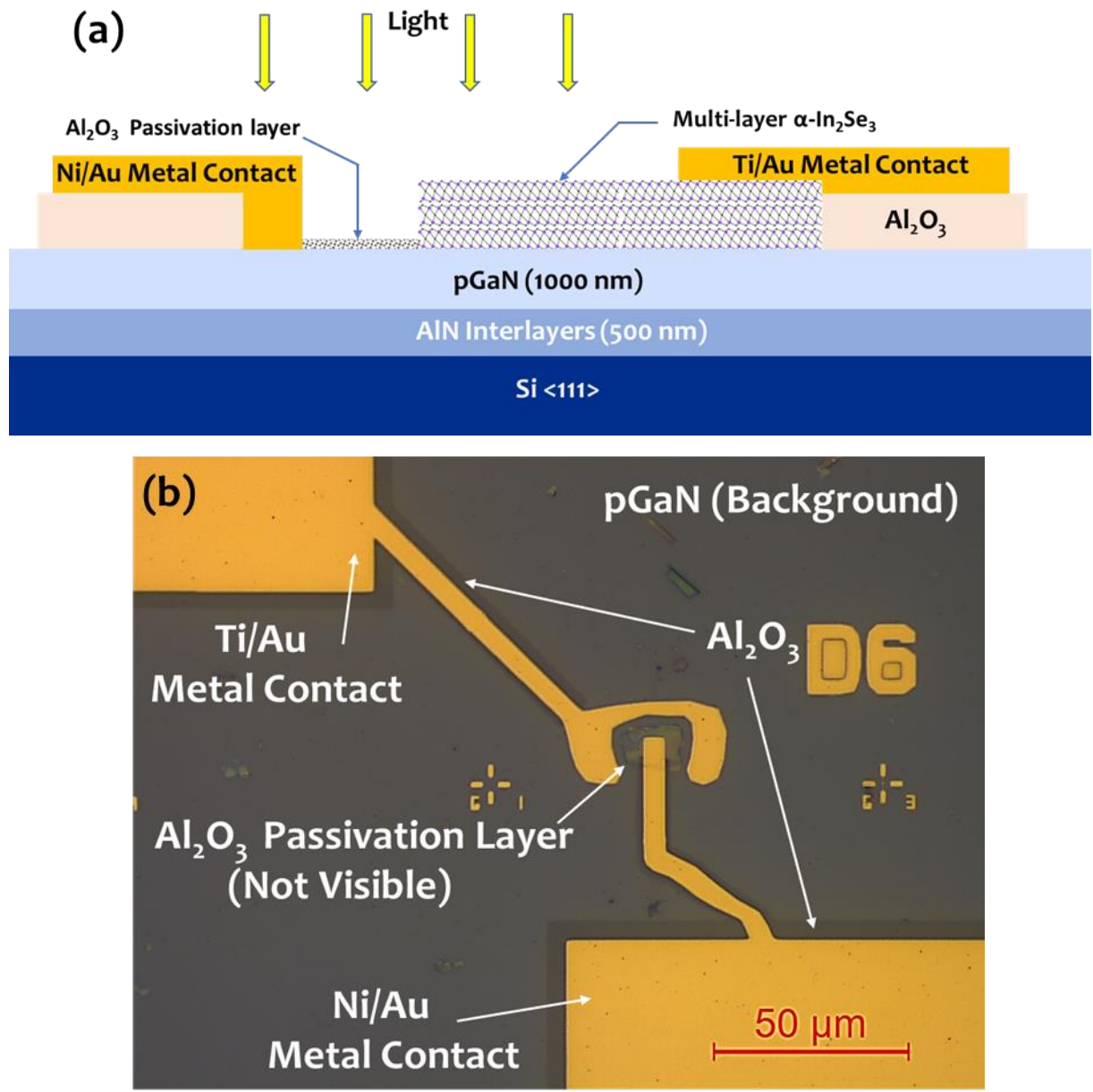

**Figure 1:** (a) Device structure for p-GaN/α-In$_2$Se$_3$ vertical heterojunction and (b) optical micrograph of as fabricated device

Fabrication of the device, with its strcture shown in Figure 1(a), was started with patterning for dielectric layer using e-beam lithography followed by deposition of 50 nm

Al$_2$O$_3$ (dielectric) using e-beam evaporation. The purpose of Al$_2$O$_3$ layer is to electrically isolate GaN underlayer from metal contact pad to prevent spurious current and allow metal contact only at intended site on both GaN and α-In$_2$Se$_3$. Next, patterning was repeated for metal deposition using e-beam lithography followed by deposition of Ti/Au (30nm/100nm) metal contact for α-In$_2$Se$_3$ using e-beam evaporation. Subsequently, Ni/Au (30nm/100nm) was e-beam evaporated to form contact with p-GaN. Finally, 10 nm of Al$_2$O$_3$ was e-bean evaporated using same procedure, as a passivation layer. Figure 1(b) shows the optical image of as fabricated device while, the fabrication flow of the device until Ni/Au can be referred in supplementary data, Figure S4.

## 3. Results and Discussion

Electrical measurements were performed using 4-probe station connected to an Agilent B1500 Semiconductor Parameter Analyzer. Keithley 2450 source-meter was used for biasing as well as measurement of photocurrent. The details about tools and methods used in optical parameter measurements are given in supplementary data (Figure S5).

### 3.1 Photocurrent and spectral responsivity measurements

Figure 2 shows the dark as well as photo I-V characteristics of p-GaN/α-In$_2$Se$_3$ heterojunction. As can be seen, ~0.08 nA of dark current was observed at 3 V during positive bias, while during negative bias, however, ~0.017 nA of current was obtained for the bias of -3V, indicating weak rectification. An ideality factor of ~3.3 was extracted by fitting in diode equation in the range of 0.6 V to 1.5 V. The deviation of ideality factor from its ideal value between 1-2 indicates possible presence of interface traps/states at α-In$_2$Se$_3$/p-GaN hetero-interface. [26]

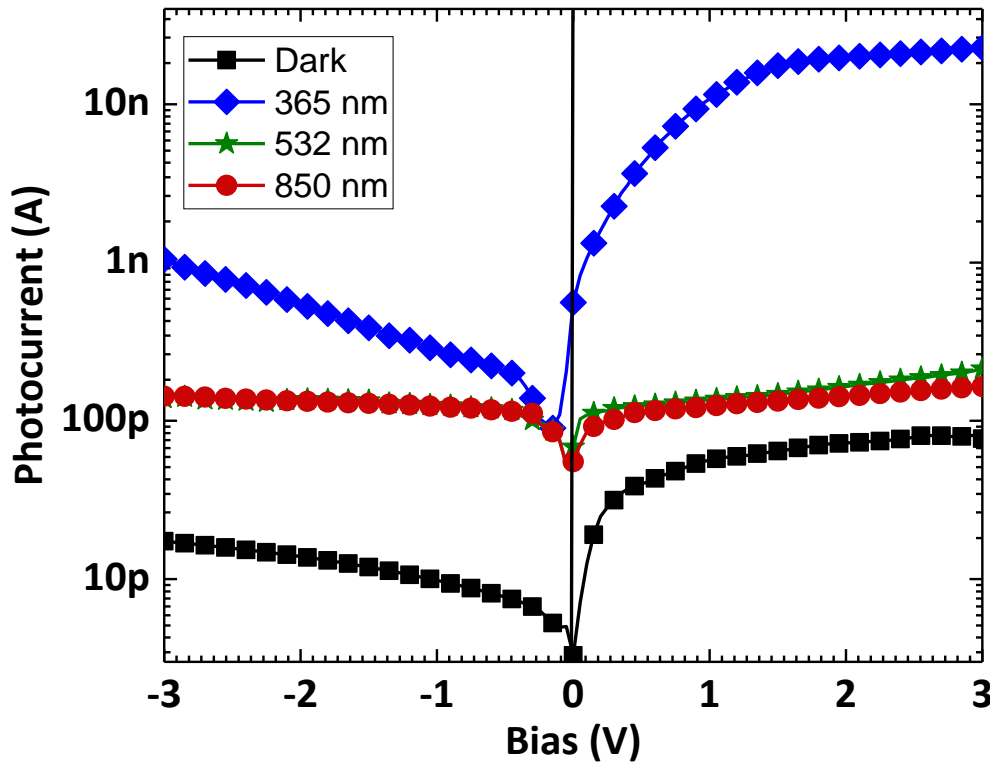

**Figure 2:** Dark and Photo I-V characteristics for p-GaN/α-In$_2$Se$_3$ heterojunction. Except at 365 nm, photocurrent showed similar trends at 532 nm and 850 nm.

Photo current measurements were then performed on the device by illuminating it at multiple wavelengths. As evident from Figure 2, at 365 nm of incident wavelength, significant difference was observed for photo currents at positive compared to negative biases. For instance, at 3 V, the photocurrent was ~10 times larger compared to that at -3V. The photo currents measured under illumination of 532 nm and 850 nm, however, were much lower and showed similar values and trends. Additionally, photo current-to-dark current ratio was found out to be ~665 in negative bias and ~310 in positive bias at 365 nm. While at 850 nm of illumination these ratios were observed to be ~75 and ~3 in that order.

Figure 3 shows the spectral responsivity (normalised with respect to active device area) graph at 3 V and -3 V. A sharp difference in responsivity (~1.9 A/W at 3V and ~0.088 A/W at -3V) was observed at the band edge of GaN which is 365 nm. After 365 nm, responsivity exhibited continuous fall until 500 nm after which a steady rise in SR value was observed until 850 nm. At 850 nm (~0.07 A/W at -3V and ~0.05 A/W at 3V), another steep fall was observed corresponding to the bandgap of α-In$_2$Se$_3$. [27][28][29]

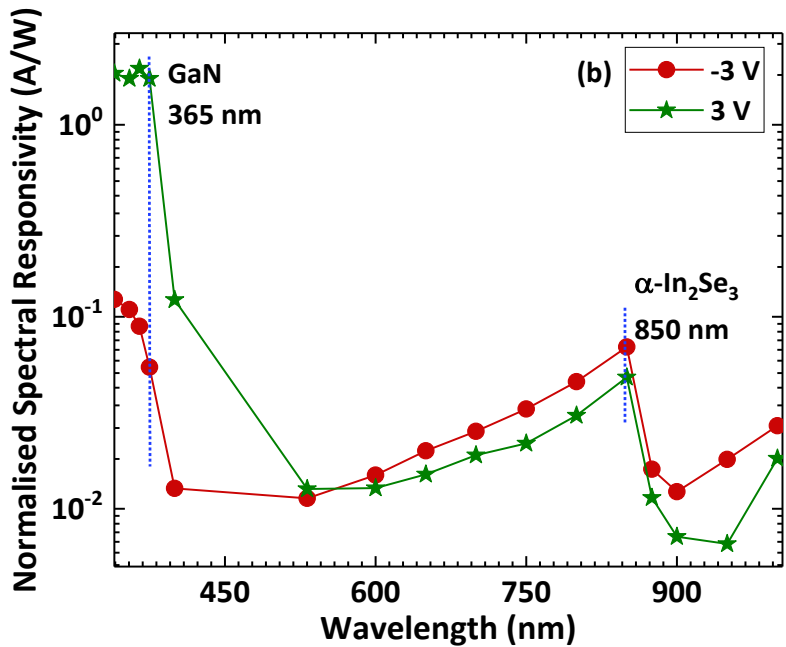

**Figure 3:** Normalised spectral responsivity graph corresponding to 3 V and -3 V biases.

It is noteworthy that responsivity showed similar trends in both positive and negative biases. Device showed higher responsivity in positive bias than in negative bias for the incident wavelength of 365 nm. At 850 nm, device showed marginally better performance at negative bias compared to positive bias. For negative bias, corresponding responsivity values at 365 nm and 850 nm were almost similar, in the range of ~70 mA/W to 90 mA/W and had a nearly 10x suppression of visible wavelengths between 400-600 nm, approaching a behaviour of band-stop filter. This shows the promise of the device for UV and near-IR photo detection.

**3.2 Current transport and photodetection mechanism.**

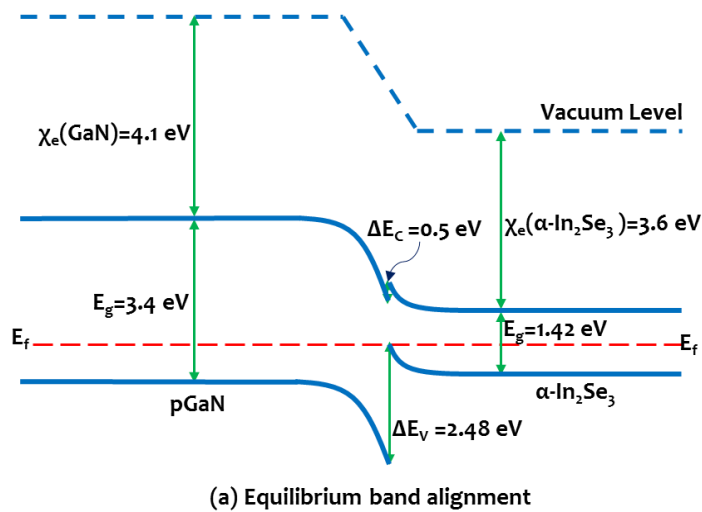

(a) Equilibrium band alignment

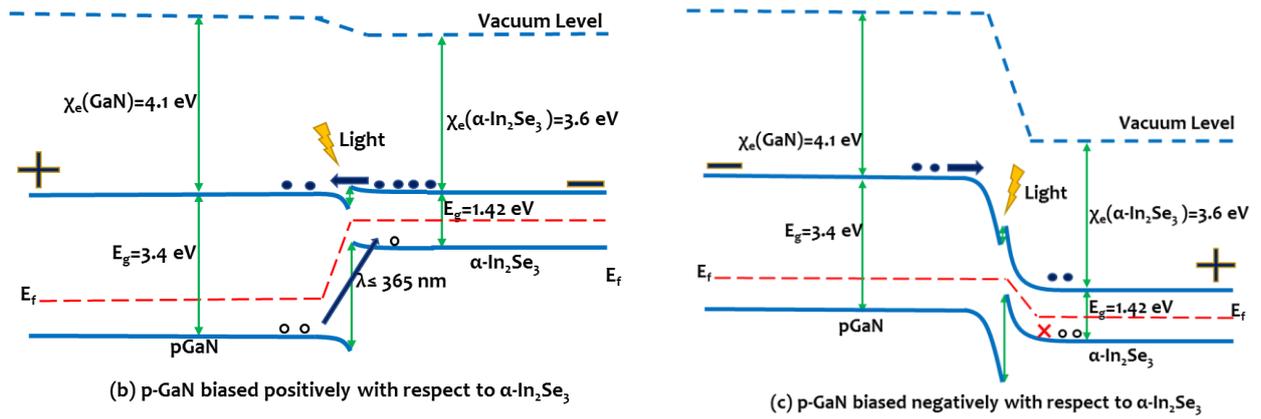

**Figure 4:** Band alignment for p-GaN/α-In$_2$Se$_3$ heterojunction photodetector in (a) Equilibrium (b) Positive bias (c) Negative bias.

In heterojunctions, band alignment plays a crucial role in photo carrier transport and detection because it can be exploited to enable efficient and selective carrier collection. α phase of In$_2$Se$_3$ has nearly same electron affinity as that of β phase (~3.6 eV).[11][27][28] Figure 4 (a) shows the equilibrium band alignment for p-GaN/α-In$_2$Se$_3$ heterojunction.

From the band alignment based on reported electron affinities, conduction band and valence band offsets of 0.5 eV and 2.4 eV respectively, between p-GaN and α-In$_2$Se$_3$ is expected. This indicates that the barrier for electron transport would be much less to overcome than that for holes, indicating dominance of electrons transport. So, when excited with incident photons, the photocurrent transport can be explained with the help of band diagrams under non equilibrium conditions. [26][30]

During positive bias, the quasi-Fermi level of GaN would shift below the quasi-Fermi level of α-In$_2$Se$_3$ (Figure 4 (b)), resulting in reduced barrier for both electrons and holes. Further, when excited with the photon of energy, hν ≥ 3.4 eV (corresponding to bandgap of GaN) from the top, both p-GaN and α-In$_2$Se$_3$, will take part in photocarriers generation. As a result, electrons and holes would flow towards their respective electrodes, contributing to the photocurrent. At 532 nm and 850 nm however, only α-In$_2$Se$_3$ would take part in photocarrier generation, leading to, a considerable reduction in photocurrent. Thus, as shown in Figure 3, keeping in mind that flake size is very small, the magnitude of photocurrent at 532 nm and 850 nm are observed to be almost similar.

During negative bias, the quasi-Fermi level of α-In$_2$Se$_3$ would be at a level lower than the quasi-Fermi level of GaN. As shown in figure 4 (c), electrons would not face any barrier, but holes, generated in Indium Selenide would not be able to cross the junction because of the large ΔEv. When excited with photon of energy, hv ≥ 3.4 eV, photogenerated electrons (minority carriers) in p-GaN would able to cross the barrier and reach other end but holes from α-In$_2$Se$_3$ will not able to reach p-GaN side. From Figure 3, we observe that, though reduced, we still got noticeable photo current at 365 nm in negative bias condition. At excitation of 532 nm and 850 nm, photocurrent was observed to be at similar level, as it was observed during positive bias, the reason behind which is still needs to be understood.

### 3.3 Performance Parameters

External Quantum Efficiency (EQE) and gain (G) are two important performance parameters for photo detector. The EQE, $\eta_e$, of a photodetector is defined as the ratio of the number of electron-hole pairts extracted from the device to the number of photons incident. If external quantum efficiency exceeds 1, then photodetector is said to have a gain. The relationship of EQE with spectral responsivity is given as [31][32] [33]

$$\eta_e = \frac{R}{R_{ideal}} = \frac{R}{(\frac{q\lambda}{hc})} \qquad (1)$$

where, (qλ/hc) is ideal or theoretical responsivity, $R_{ideal}$ and R is measured responsivity and $\eta_e$ is EQE.

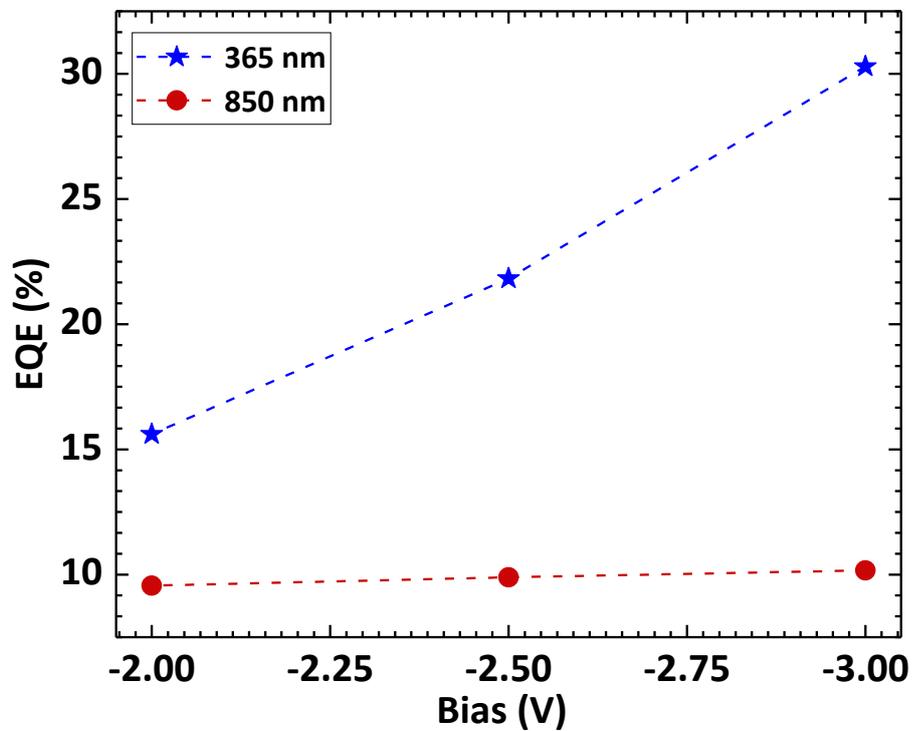

**Figure 5:** Quantum Efficiency as a function of applied bias for p-GaN/α-In$_2$Se$_3$ heterojunction.

Figure 5 shows measured EQE as a function of applied bias. It can be observed that, EQE showed increasing trend with applied bias for illumination at 365 nm, indicating transit time limited performance.[34][35][36] For 365 nm, EQE was calculated to be ~30% at the applied bias of -3 V. Whereas for illumination at 850 nm, EQE remained almost constant around the value of ~10%. In other words, device did not show any internal gain, since calculated responsivity values of ~90 mA/W, at 365 nm and ~70 mA/W at 850 nm, were lower than theoretical responsivity values, 0.293 and 0.684 respectively. After this, power-dependent responsivity analysis was performed on the device at 365 nm of incident wavelength. QE tool was used as power source. Built-in neutral density (ND) filters of the QE tool were used to change the incident power intensity.

Figure 6 shows the normalised spectral responsivity as a function of normalised incident power at 365 nm. The device exhibited **$P^{\sim -0.4}$** dependence on incident power at 365nm of illumination. The term β in the power expression $R \sim P^\beta$ is indicative of the linearity of the photodetector over the range. As β→0, device is said to be more linear. However, in presence of traps which cause the photo-induced barrier lowering associated with gain in these films, the photocurrent tends to get saturated at higher powers, leading to a sub-linearity and deviation from the ideally expected value of β=0. A larger deviation from 0 implies larger saturation of gain mediated trap states at increased powers and a slow release of trapped photo-generated carriers, thus responsivity no more remains independent of incident power. As responsivity saturates at higher incident power, indicating that it is indeed limited because of saturation of photocurrent at higher powers due to traps and impurities present in GaN and heterointerface.[11][36][37] At higher incident power traps capture generated excess photo carriers (electrons or holes depending upon nature of traps). There would always be availability of enough excess generated photo carriers to keep these traps in filled position, causing reduction in number of photogenerated carriers. This leads to reduction in net photo current, causing reduction in the ratio of net photocurrent to incident optical power ($I_p/W_i$) hence the responsivity value. At lower incident power levels, number of the net photogenerated carriers is low, which is further reduced because of traps. But there is always possibility that photocarrier can escape from traps or can be released by trap due to applied bias or incident optical power. Thus, net photocurrent does not reduce below certain level maintaining the $I_p/W_i$ ratio. Additionally, low incident power also contributes to the increase in the $I_p/W_i$ ratio. As can be seen from Figure 6, exponent value of ~-0.45 indicates fairly linear device, indicating lower density of traps/impurities at heterointerface or in layered material.

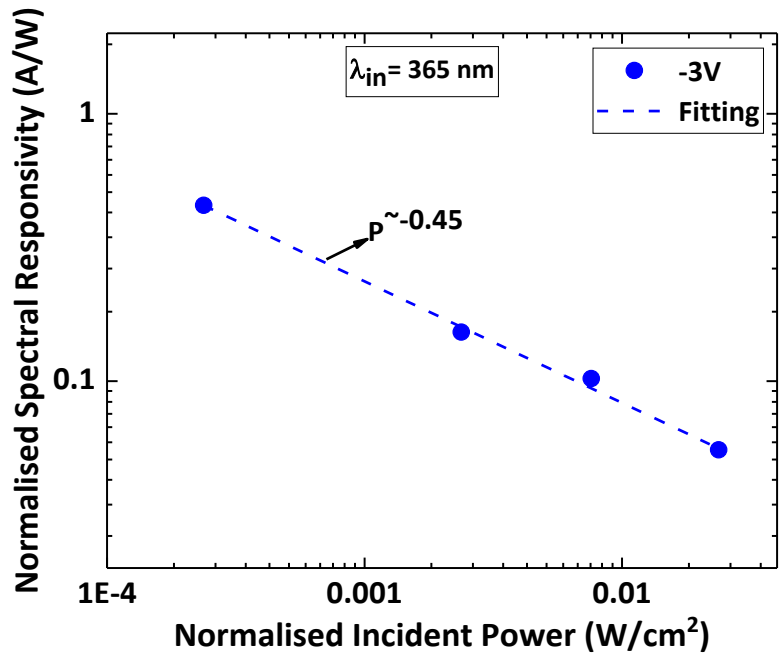

**Figure 6:** Normalised responsivity as a function of normalised incident power at 365 nm for p-GaN/α-In$_2$Se$_3$ device.

Transient response is another critical figure of merit for photodetectors. Device was tested for transient response at 365 nm and 850 nm. A sharp rise/fall in photocurrent was observed when light was turned ON and OFF respectively (Figure 7). At 365 nm, the rise time was measured to be ~0.2 s while the fall time was measured to be ~0.13 s. At 850 nm, photocurrent measurement was affected because of noise. So current profiles were smoothened using mathematical smoothing tool. At 850 nm, rise time was observed to be around ~0.26 s and fall time was found to be in the range of ~0.42 s. No persistent photoconductivity (PCC) was observed, leading to improvement in speed. Transient response was marginally slower under 850 nm illumination compared to at 365 nm. This indicates small trap/dislocation densities inside the semiconductors or at heterointerface.

Specific detectivity (D*) is one of the important performance parameters of photodetectors. Specific detectivity is measure of the capability of photodetector to detect incident photon in high dark current environment or/and in noise background. A higher detectivity value indicates better photodetector. D* values for the devices under study were calculated for given incident power using the following equation, [32] [33] [34]

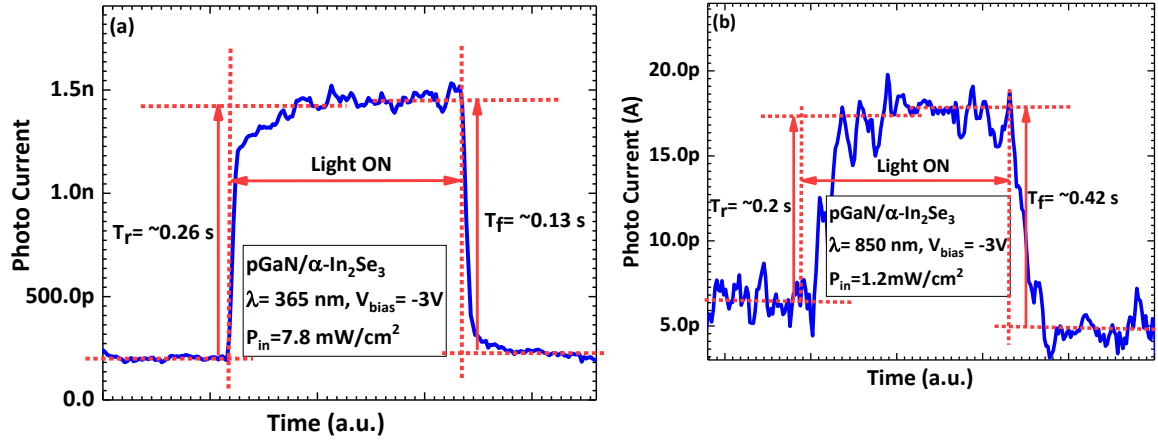

**Figure 7:** Shows transient graphs obtained at (a) 365 nm and at (b) 850 nm for bias of -3 V.

$$D^* = q\eta_e/h\nu \left[\frac{4\kappa T}{R_o A} + \frac{2 q I_{Dark}}{A}\right]^{-\frac{1}{2}} \quad Jones \quad (2)$$

Where, $R_o$ is resistance at zero bias, $I_{Dark}$ is dark current, A is active area of the device and q is the charge of an electron, $\eta_e$ is EQE and other symbols have their usual meanings.

D* was calculated to be 4.6 x $10^{10}$ Jones at -3 V at 365 nm of illumination. For 850 nm of incident wavelength, D* was found out to be 3.6 x $10^{10}$ Jones at -3 V. This is first report of D* at 850 nm which is purely contributed by α-In$_2$Se$_3$. Given the low value of photocurrents, D* as high as in the range of ~$10^{10}$ Jones shows the promising nature of such architectures, which if improved, will yield better performance parameters.

It would be noteworthy to compare the performance of this device with the performance of our previously reported β-In$_2$Se$_3$/GaN dual-band photodetector, [38] though later was fabricated in MSM architecture. The p-GaN/α-In$_2$Se$_3$ device clearly showed the promise for band-stop filter with ~10x suppression in visible spectrum of light against β-In$_2$Se$_3$/GaN, which could not suppress the detection in the same range. The normalised spectral responsivity value was observed to be doubled at 850 nm (~70 mA/W) at much lower bias of -3 V compared to the value of ~32.7 mA/W at an applied bias of -20 V in previous work. Additionally, photo-to-dark current ratio at 850 nm was improved to ~75 from ~5 in that order. A significant improvement was also observed in speed of the device with no persistent photoconductivity. Table 1 shows the comparison of performance parameters of reported device with our previous work as well as with other state-of-the-art devices with similar architecture. It is to be noted that, while laser us used as a light source in most of the works, our device shows better performance using broad band light source with significantly lower incident power.

| Sr. No | Device | Performance Parameters Responsivity (R) @ λ, Detectivity | Reference |
|---|---|---|---|
| 1 | p-GaN/α-In$_2$Se$_3$ vertical heterojunction | R= 88 mA/W @ 365 nm<br>R= 70 mA/W @ 850 nm<br>D*= ~3.6 x 10$^{10}$ @ 850 nm | This Work |
| 2 | MoS$_2$/p-GaN heterojunction | R= 50 A/W @ 532 nm,<br>R= 300 A/W @ 405 nm,<br>Both values at 0.1 W/cm$^2$ of laser | [9] |
| 3 | β-In$_2$Se$_3$/GaN MSM | R= 1.27 A/W @ 365 nm<br>R= 32.7 mA/W @ 850 nm<br>D*= ~1.6 x 10$^9$ Jones @ 805 nm | [38] |
| 4 | MoS$_2$/GaN MSM | R= 127 A/W @ 365 nm<br>R= 33 A/W @ 685 nm<br>D*= ~3.3 x 10$^{11}$ Jones @ 532 nm | [38] |
| 5 | MoS$_2$-GaN heterojunction | R= 1 A/W @ 405 nm, 2mW<br>R= 100 mA/W @ 638 nm, 2mW<br>D*= ~10$^{10}$-10$^{12}$ Jones, in the range of 405 nm to 650 nm | [39] |
| 6 | p-Si/γ-In$_2$Se$_3$ heterojunction | R= 0.54 A/W @ 365 nm<br>D*= 3.5 x 10$^{12}$ Jones @ 365 nm | [40] |
| 7 | MoS$_2$/PbS heterojunction | R= 2 x 10$^5$ A/W @ 600 nm<br>R= 1 x 10$^5$ A/W @ 1000 nm<br>D*= 5 x 10$^{11}$ Jones @ 635 nm | [41] |

**Table 1:** Summary of the performance parameters in this work along with other reported dual band structures for comparison purpose.

## 4.0 Effect of annealing on device performance

In many applications, post fabrication, devices are subjected to thermal annealing, which helps in improving the metal-semiconductor contact, in mitigating the effects of dangling bonds on the surface, reducing the number of traps particularly at dielectric/semiconductor interface.[42][43][44] In many cases however, annealing also changes physical or chemical properties of the materials causing either shift or complete change in the behavior of the device. The change can either be desirable or undesirable.

Annealing of p-GaN/α-In$_2$Se$_3$ device was done in AS-ONE Rapid Thermal Annealing system at 300 °C for 2 minutes in N2 ambient. Devices were then again investigated for their performance.

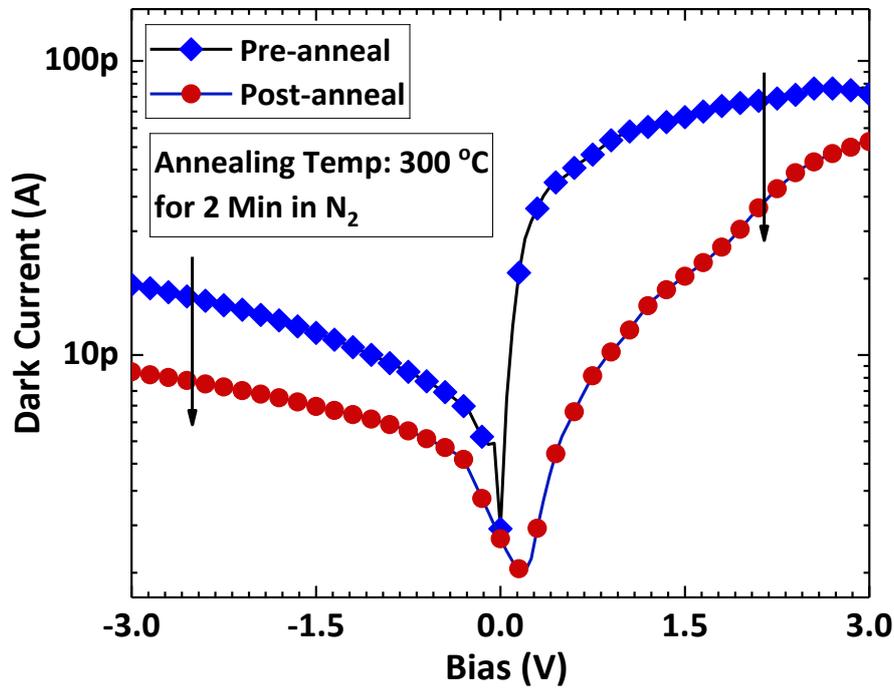

**Figure 8:** The dark I-V of the device after annealing at 300 °C in N$_2$ ambient. A decrease in dark current was observed.

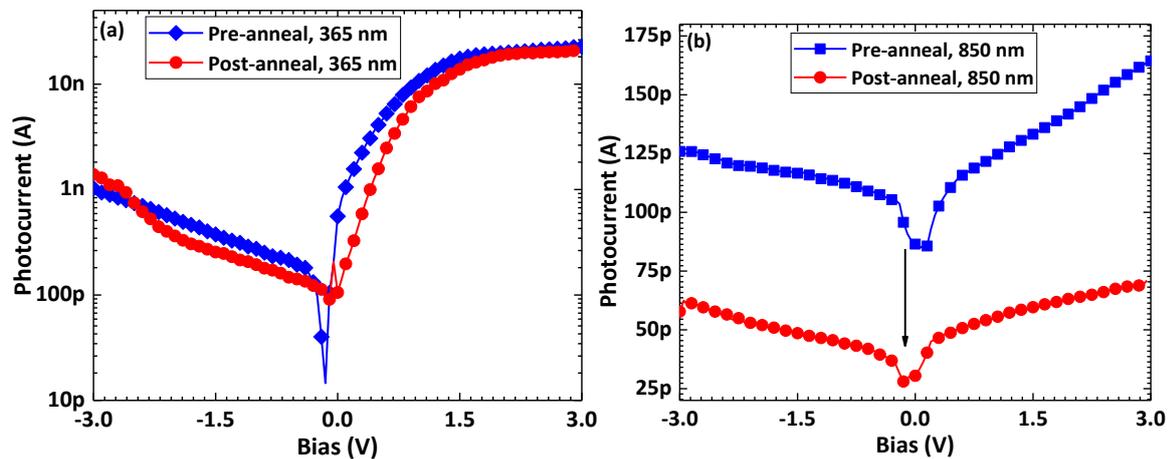

**Figure 9:** (a) and (b) show the change in photocurrent at 365 nm and 850 nm respectively for both pre and post annealing conditions.

Figure 8 shows the comparative graph of dark current performance of the device before and after annealing. A clear decrease in the dark current was observed. Reduction in dark current indicates better performance in the presence of shot noise. Figure 9 (a) and (b) show the pre and post-annealing photocurrent characteristics of the device at 365 nm and 850 nm of illumination. Figure 9 (a) clearly shows that photocurrent corresponding to 365 nm illumination remained almost unchanged after annealing the sample. GaN didn't undergo any change at the annealing temperature as expected However, appreciable reduction in photocurrent was observed at 850 nm with current falling by nearly 2 orders of magnitude compared to pre-anneal data. Had there been a reduction in traps or dangling bonds in Indium Selenide or an improvement in the hetero-interface after annealing, the photocurrent would at least be expected to remain unchanged if not higher in magnitude. The reduction in photocurrent indicates a change in physical or material of Indium Selenide.

The spectral responsivity of the device remained largely unchanged as shown in Figure 10, except for a reduction in the responsivity value at 850 nm corresponding to both positive and negative bias. This can be attributed to the reduction in photocurrent at 850 nm as we had seen in Figure 9 (b).

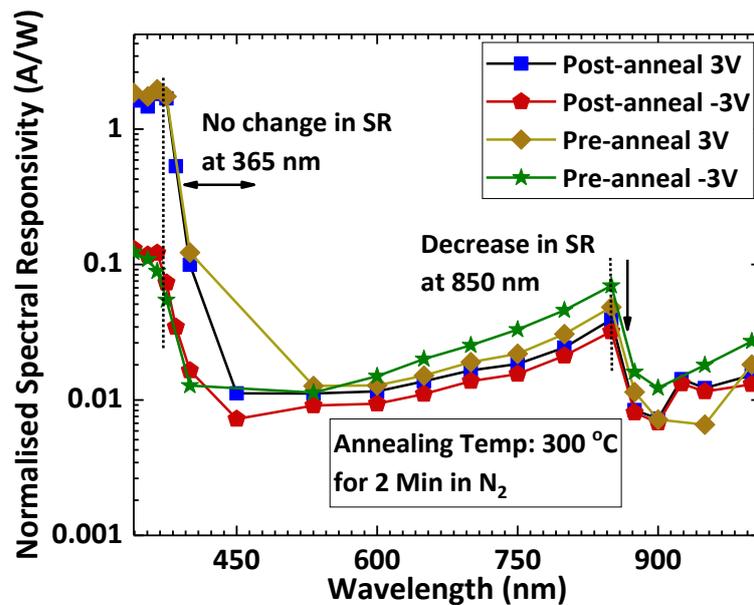

**Figure 10:** Pre and post annealing SR graph for the device. Device showed reduction in Sr at 850 nm but no change at 365 nm.

Raman scan was performed on sample to investigate the possible changes in performance parameters. As anticipated, phase of $In_2Se_3$ changed from α to β when annealed at 300 °C in $N_2$ ambient. Figure 11 shows the comparative Raman spectra for $In_2Se_3$ in both pre and post annealing conditions. The Raman peaks corresponding to the vibrational modes E and A in α phase of $In_2Se_3$ are clearly visible at 88.2, 104.2, 180.9 and 192.6. Post anneal Raman spectra clearly shows change in peak positions, corresponding to vibration mode 'A' in β phase of $In_2Se_3$, at the positions of 110.0, 175.0 and 205.0.

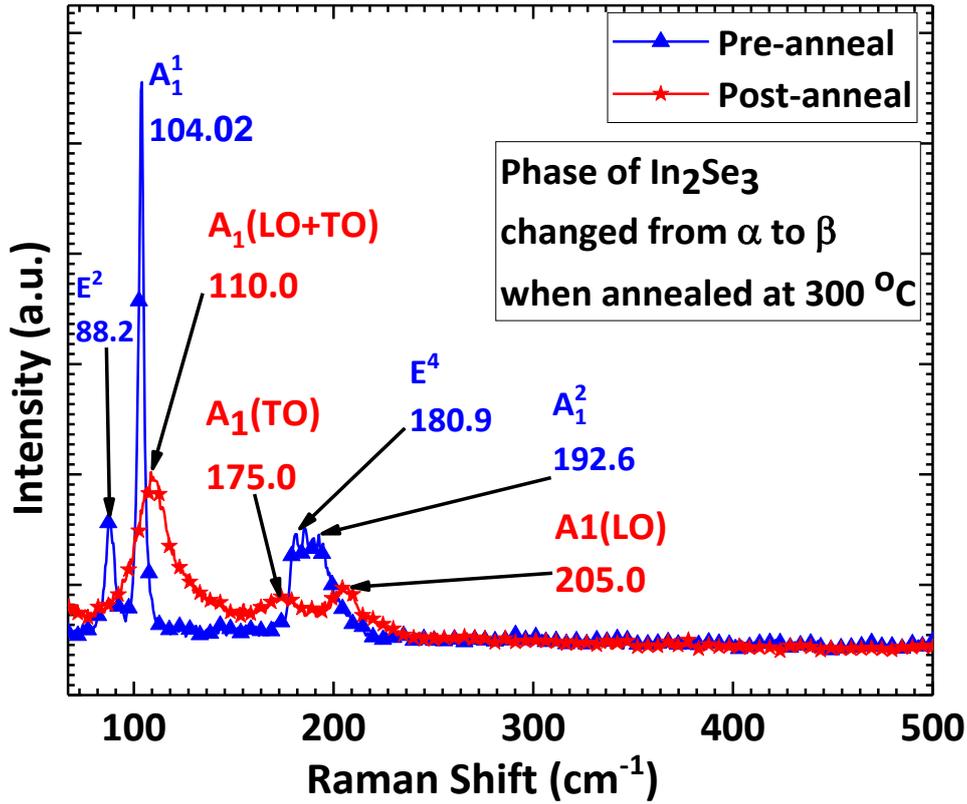

**Figure 11:** Raman spectra obtained for $In_2Se_3$ before and after annealing showing change in phase from α to β.

The change from α to β in $In_2Se_3$ indicates the change in crystal structure.[16][45] Both the phases belong to rhombohedral crystal structure but different space group (R3m and R$\bar{3}$m respectively). In β phase, atoms in primitive cell occupy smaller volume than in α phase. A few reports suggest that electrical resistance in β phase is approximately 2 to 3 orders of magnitude less than in α phase,[45][46] but in our case, we did not observe reduction in resistance (increase in current). Instead we observed significant reduction in photocurrent at 850 nm as well as in dark current. This suggests that when the phase changed from α to β, material quality of $In_2Se_3$ degraded significantly. Note that, in all available reports, phase change of $In_2Se_3$ from α to β or growth of β phase is done in highly controlled environment with proper parameter monitoring.[47][48] Phase transition from α to β has been shown to make $In_2Se_3$ amorphous also.[49] We believe that, this could be the reason for the fall in dark current as well as photocurrent at 850 nm for the device reported in this study. An amorphous β-$In_2Se_3$ will certainly scatter the charge carriers and will reduce the dark current also. When excited with light, photogenerated carriers would experience an increase in scattering thus leading to reduction in the photocurrent and responsivity values.

## 5.0 Conclusions

In conclusion, we demonstrated dual wavelength photodetection using p-GaN/α-$In_2Se_3$ vertical heterojunction, with distinct peaks at 365 nm and 850 nm and with good responsivity and transient responses. Device showed better promise for UV band detection under 'positive bias' and NIR band detection in 'negative bias' conditions thus proving useful under both the bias conditions. Device also exhibited good response time showing promise in the application where faster switching is required. Additionally, we observed that when device was annealed at 300 °C in $N_2$ ambient for two minutes, because of the phase change of $In_2Se_3$ from α to β and degradation in quality, significant reduction in photocurrent was observed at 850 nm.

**Acknowledgements**

We acknowledge funding support from Ministry of Human Resource Development (MHRD), Government of India (GOI) through NIEIN project as well as from Ministry of Electronics and Information Technology (MeitY), GOI and Department of Science and Technology (DST), GOI through NNetRA. This work is funded under DST/SERB Grant, No. 01482, MeitY, GOI.

The authors would like to acknowledge support from the Research and Development work undertaken under the Visvesvaraya Ph.D. scheme of Media Lab Asia, Ministry of Electronics and Information Technology (MeitY), Government of India. Authors would also like to thank Micro Nano Characterization Facility (MNCF) and National Nano Fabrication Centre (NNFC) staff at CeNSE, IISc for their kind support for carrying out this work.

# UV/Near-IR dual band photodetector based on p-GaN/α-In$_2$Se$_3$ heterojunction

*Swanand V. Solanke[1*], Rohith Soman[1,2], Muralidharan Rangarajan[1], Srinivasan Raghavan[1] and Digbijoy N. Nath[1]*

[1]Centre For Nano Science and Engineering (CeNSE), Indian Institute of Science, Bengaluru, Karnataka -5600129 (INDIA)

[2]Depart of Electronic and Communication Engineering, Indian Institute of Science, Bengaluru, Karnataka -560012 (INDIA)

Correspondence email: swanands@iisc.ac.in

1. **XRD analysis for p-GaN**

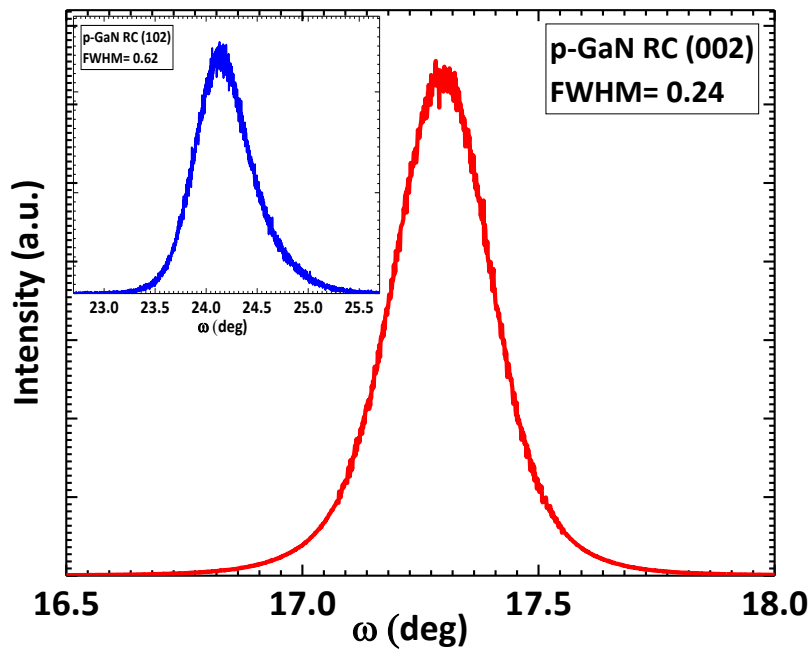

**Figure S1:** XRD (002) and (102) plane Rocking Curves for p-GaN grown for our work.

2. **AFM and Raman analysis for confirming multilayer nature and α phase of In₂Se₃ flake.**

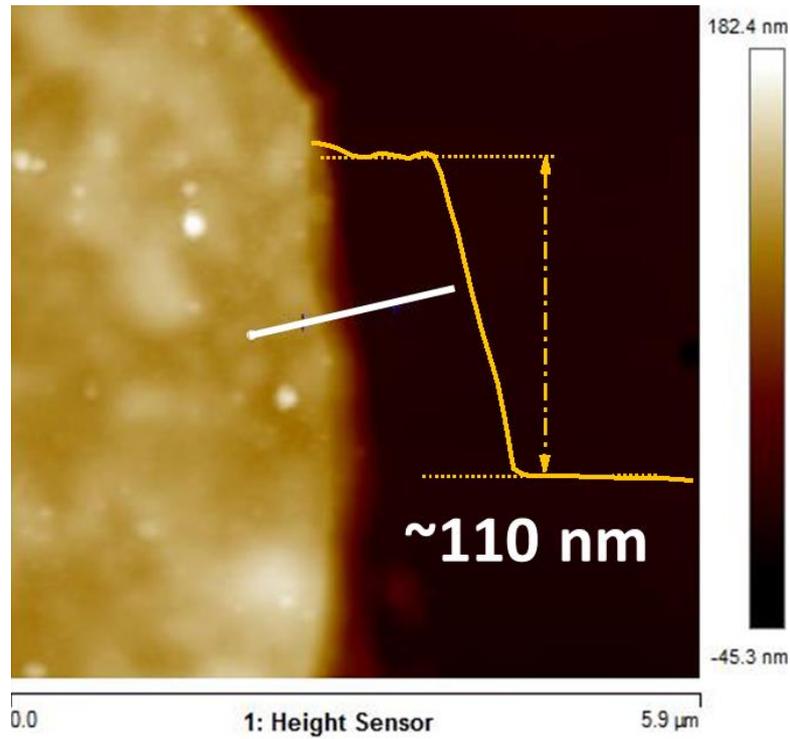

**Figure S2:** AFM analysis for α-In₂Se₃ confirming multilayer nature of the flake.

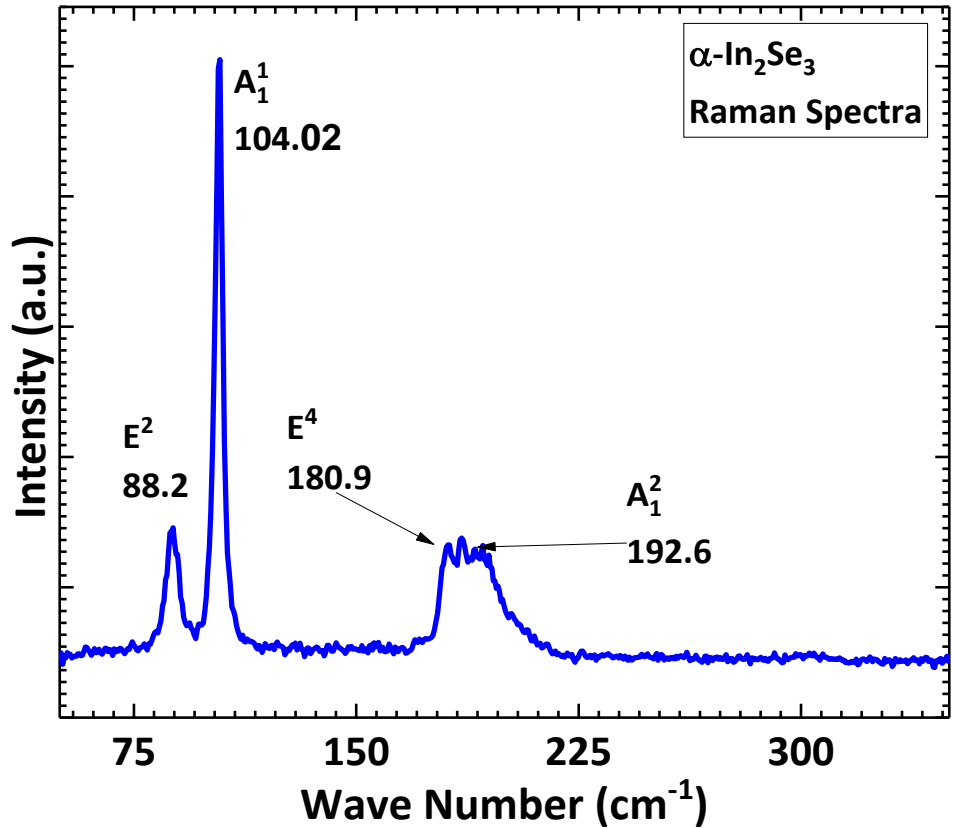

**Figure S3:** In2Se3 Raman Spectra confirming the α phase.

3. **Fabrication flow**

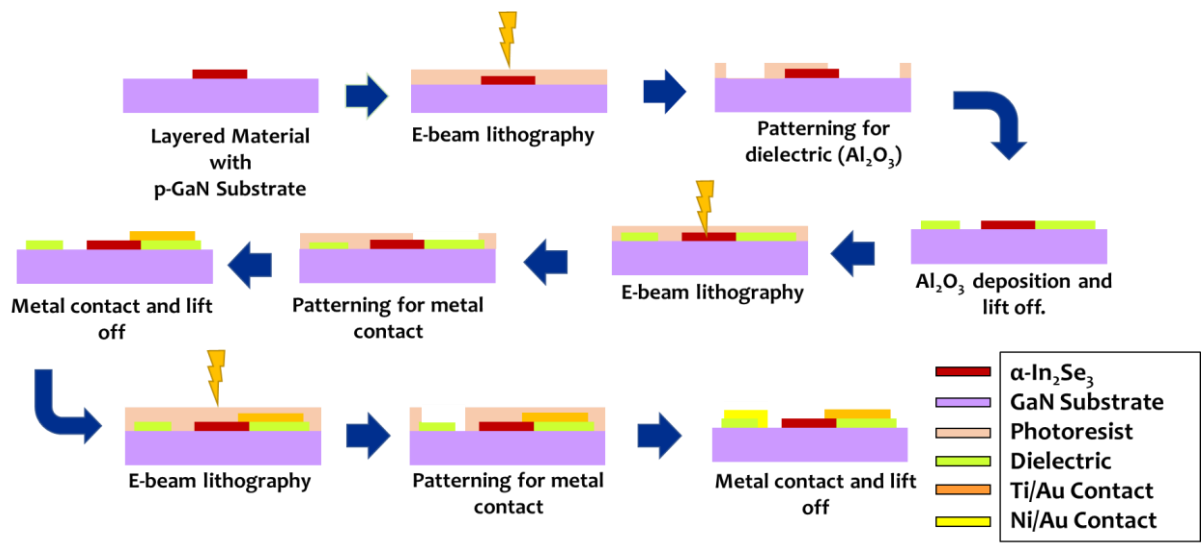

Figure S4: Fabrication flow for p-GaN/α-In2Se3 photodetector until the step of Ni/Au metal deposition. For $Al_2O_3$ passivation layer, lithography and deposition procedure was same as $Al_2O_3$ isolation layer deposition.

4. **Methodology and tool for optical measurements.**

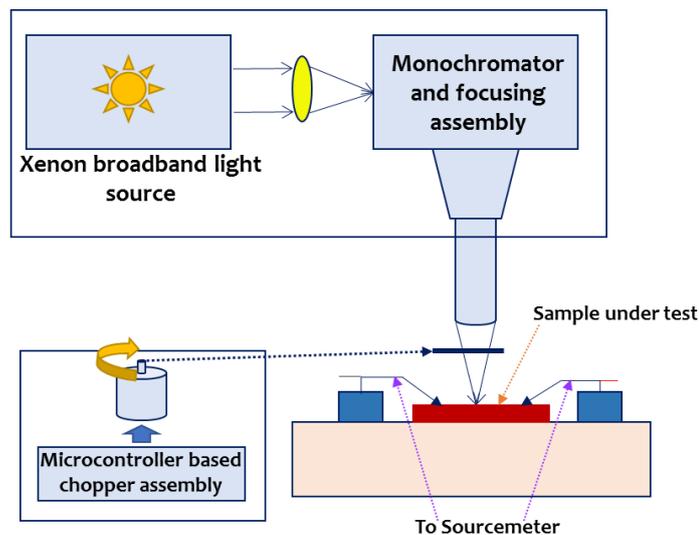

**Figure S5:** Schematic showing experimental setup for optical I-V and other related measurements.

ScienceTech Quantum Efficiency tool was used for photo I-V measurements. Tool consists of 150 W Xenon broadband light source. Light coming from the source was made to fall on sample under test using focusing-grating-monochromator-filter assembly. For transient measurements, microcontroller-based chopper assembly was used. Keithley 2450 source-meter was used for biasing as well as measurements. The optical spectrum range of light output of the tool is 230 nm to 1100 nm.